# GNSS Spoofing Detection by Crowdsourcing Double Differential Pseudorange Spatial Distribution

Xin Chen, *Senior Member, IEEE*, and Kai Wang

*Abstract*—It is widely known that spoofing is a major threat that adversely impacts the reliability and accuracy of GNSS applications. Current spoofing detection methods mostly work with advanced signal processing in a receiver. As more and more personal devices contain GNSS chipsets, spoofing detection by crowdsourcing receivers has become an attractive way to detect and localize spoofers. In this paper, a crowdsourcing double differential pseudorange spatial (D$^2$SP) random set is constructed and the distribution of the set is derived. Based on the variance of the D$^2$SP set, a tri-level hypothesis detection algorithm is designed to classify spoofing-free, fully-spoofed, and partially-spoofed cases in the region of interest (ROI). It does not require the prior knowledge of the truth positions or relative distances of the receivers. Simulation test results show that the proposed D$^2$SP spoofing detection method has the advantages of lower computational complexity and higher tolerance for multipath errors compared with the generalized likelihood ratio test (GLRT) method that is the current mainstream spoofing detection algorithm based on multiple receivers' differential pseudoranges. Moreover, it also shows better flexibility for different sizes of ROI and numbers of the crowdsourcing receivers.

## I. INTRODUCTION

As Global Navigation Satellite Systems (GNSS) have been widely utilized in a variety of fields, spoofing is becoming a major threat that adversely impacts the reliability and accuracy of GNSS applications. Victim receivers would be misled by the counterfeit GNSS signals broadcast by a spoofer to deduce false position and/or time fixes [1].

Spoofers can be classified as meaconers, signal simulators, or receiver-repeaters based on the spoofing signal generation method [2-4]. A meaconer, also known as a repeater, simply broadcast amplified received authentic GNSS signals. A GNSS simulator-based spoofer transmits simulated GNSS signals at a power level greater than authentic signals. A receiver-repeater, also referred to as an estimate-and-replayer, reconstruct spoofing signals based on the received authentic signals by manipulating transmit times, Doppler shifts, and even navigation messages. This type of spoofer can launch an approach-and-drag-off attack that deceives a commercial receiver [3,4].

Traditionally, spoofing countermeasures are relying on receiver-side signal processing strategies. The signal measurements, such as carrier-to-noise ratio (C/N$_0$), automatic gain control (AGC) value, post-correlation value, etc., are used as signal quality metrics for spoofing monitoring [5,6,8]. Multi-antenna methods are adopted for spoofing discrimination and mitigation [18,19].

As location-based services (LBS) are broadly used in various civil applications, GNSS chipsets have been embedded in numerous devices, such as personal mobile phones, land or aerial vehicles, communication or power grid timing stations, etc. The Android 7.0 version and above have enable the access to GNSS signal metrics including C/N$_0$, Doppler frequency, pseudorange, etc [41]. The large number and the wide spread of the receivers lead to the crowdsourcing method as an attractive approach for GNSS jammer monitoring and localization in a relatively large area [30-32], because it avoids the requirement of deploying dedicated devices to monitor the interference, thus greatly reducing the deployment costs of the system. The representative ones are the mobile phone raw measurements integrity method [33], the cross-checking between GNSS and cellular network measurements [34], and the generalized likelihood-ratio test algorithm based on multiple receivers' double differential pseudoranges (DP-GLRT) [39]. The crowdsourcing method can help authority organizations to find interference more efficiently and protect the GNSS usage for the public from spoofing harms,

This paper presents a new spoofing detection method by exploiting the spatial diversity of the double differential pseudoranges of crowdsourcing receivers, which is named as double differential pseudorange spatial distribution (D$^2$PS). A hypothesis detection is developed based on the D$^2$PS distribution to classify different spoofing scenarios. The proposed method offers several advantages over the state-of-the-art ones:
- First, it only uses the pseudorange measurements and the

Xin Chen is at the School of Electronic Information and Electrical Engineering, Shanghai Jiao Tong University, Shanghai, China. (email: xin.chen@sjtu.edu.cn).

Kai Wang is at the Beijing Institute of Tracking and Telecommunication Technology, Beijing, China

reported positions of the receivers. There are no needs of the knowledge about the real relative position relationship of those receivers or the support of network timing\location information.

• Second, it can properly distinguish not only the case of fully-spoofing but also the case of partially-spoofing that the DP-GLRT method will have difficulties to cope with.

• Third, the proposed method has more tolerance for multipath errors due to the proper use of the spatial location diversity of the crowdsourcing receivers.

The remainder of this paper is organized as follows: Recent related works on GNSS spoofing detection are firstly reviewed in Section II. Then, the mathematic model for spoofing effects on the crowdsource receivers in a monitoring area is presented in Section III. In Section IV, the distribution of the proposed D²PS samples under various spoofing scenarios, as well as the corresponding variance model, is derived. Section V presents a resizing technique to determine the proper dimension size if the receivers nonuniformly spread in the monitor area. A spoofing detection method based on the variance of the D²PS samples is developed in Section VI. In Section VII, comprehensive simulation tests are performed to evaluate the performance of the proposed D²PS method in comparison with the DP-GLRT method. Finally, Section VIII concludes the study.

## II. RELATED WORK

The usual spoofing detection ways are developed at the receiver side and with the indicators based on $C/N_0$ value [5], AGC value [6], or pre-despreading signal power [7]. The received signal correlation function and signal quality metrics could also be designed for spoofing monitoring [8,9,10,11]. Some other techniques look at the consistency among the estimated signal parameters, such as clock drift patterns [12,13], Doppler frequency or carrier phase differences [14,16], pseudorange residuals [15], Kalman innovation values [17], etc. Spatial properties of the received signals can be exploited to discriminate and mitigate spoofing attacks. These examples are multiple-antenna-based method [18,19,20], cooperative receivers at different locations [21-23], and encryption mechanisms among mutual signals or receivers [24-27]. Furthermore, complementary metrics can be jointly used to increase the probability of spoofing detection [28,29].

As to the advent of the crowdsourcing technique, GNSS spoofing detection by a mobile phone was proposed in [33-35], where the reported GNSS positions or times are cross-checked with the reported cellular network positions or times. A spoofing alarm will be raised if these two reports of positions or times are dramatically different. Subject to the accuracy of the position/time estimation by the network, this method can only detect the spoofing attack if a large position or timing error occurs. In [36], a crowdsourcing method was proposed to detect and localize GPS spoofers by leveraging aircraft messages of the automatic dependent surveillance broadcast (ADS-B) system and the Flarm system. Another widely used method is a generalized likelihood-ratio test based on multiple receivers' double differential pseudoranges (DP-GLRT) [37-40]. The principle of this method is to explore the propagation similarities of the spoofing signals between different receivers. However, the DP-GLRT method will often fail if some of the receivers are spoofed but others are not. It also has the disadvantage of vulnerability to the multipath influence in urban environments. In the following, the proposed D²PS method will be described to overcome these aforementioned problems.

## III. SPOOFING MODEL TO CROWDSOURCE SENSORS

In a spoofing scenario, the total received signal $x(t)$ by a GNSS receiver can be expressed as [2,42]:

$$x(t) = \underbrace{\sum_{i \in J_a} A_i F_i^a(t)}_{x_a(t)} + \underbrace{\sum_{q \in J_s} A_q F_q^s(t)}_{x_s(t)} + n(t), \quad (1)$$

where $x_a(t)$ denotes the authentic signals transmitted by real GNSS satellites, and $J_a$ is the authentic signal set. $x_s(t)$ denotes the counterfeit/spoofing signal emitted by spoofers, and $J_s$ is the counterfeit signal set. $n(t)$ is the thermal noise. $F_i^a(t)$ denotes the signal of the $i$th authentic satellite and $F_q^s(t)$ the signal of the $q$th counterfeit satellite. $A_i$ and $A_q$ correspond to the amplitudes of the authentic and counterfeit signals, respectively. If $J_s = J_a$, it is known as the matched spoofing attack typically associated with a meaconer or a receiver-replayer. If a GNSS simulator is used, it is possible that $J_s \neq J_a$.

When a spoofing signal impinges on nearby receiver antennas, the impact depends on the power of the spoofing signal as well as the distance between the antennas of the victim receiver and the spoofer. As indicated in [11,43], the spoofing signal power will increase the receiver's post-correlation noise floor and affect the number of acquired authentic satellites. If the equivalent spoofing to authentic signal power ratio (SAPR) at receiver's baseband circuit exceeds a certain threshold (typically 30~40 dB for a commercial receiver), the authentic signals will be overwhelmed by the enhanced cross-correlated noise floor. In this case, the victim receiver will completely lose lock to the authentic signals and re-acquire the counterfeit ones, as a consequence of which the receiver will fix and report the counterfeit positions.

According to [44], the overwhelming-spoofed range $L_{TOV}$ impacted by a spoofer with an emitting power $P_T^s$ is estimated as

$$L \leq L_{TOV} = \frac{\lambda}{4\pi} 10^{\frac{P_T^s + G_T^s + G_R^s - P^a - G_R^a - Ls_{env} - SAPR_{TOV}}{20}}, \quad (2)$$

where $G_T^s$ is the spoofer's transmit antenna gain, $G_R^a$ and $G_R^s$ are the gains of receiver's antenna for authentic and spoofing signals, respectively, $\lambda$ is the wavelength of the radio signal, and $Ls_{env}$ is the additional propagation loss of the spoofing signal through the terrestrial environment, such as urban. $G_R^a$ and $G_R^s$ are typically different because authentic signals normally arrive from high elevations, whereas spoofing signals typically arrive from low or even negative elevations.

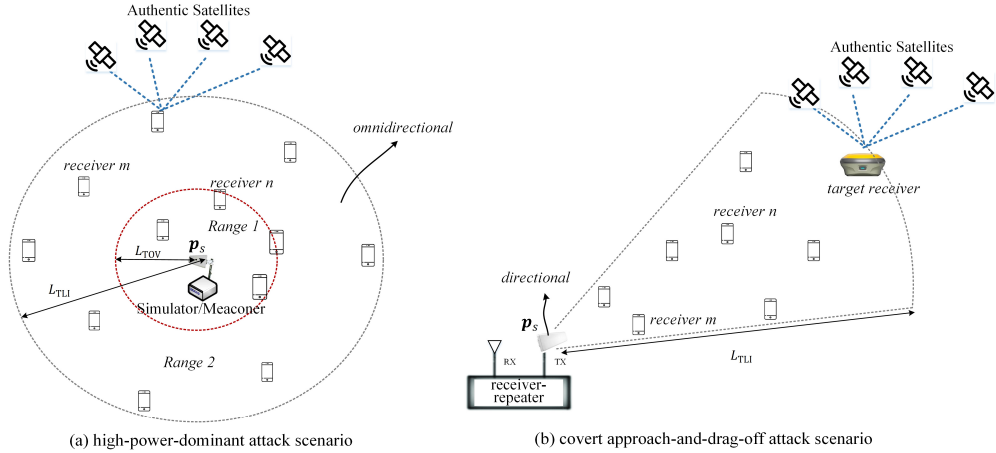

Fig.1 Two illustrative examples of the spoofing scenario. (a) The high-power-dominant attack scenario: It assumes that the spoofer antenna is omnidirectional, $G_T^s = 0$ dB, $P_T^s = 0$ dBm, $G_R^s = -10$ dB, $Ls_{env}=5$ dB, $P^a = -128$ dBm, and $G_R^a = 0$ dB. Then, the overwhelming-spoofed range $L_{TOV}$ (red dashed circle) and the risky-spoofed range $L_{TRI}$ (grey dashed circle) are, respectively, 120 m and 3800 m. (b) The stealthy approach-and-drag-off attack scenario: It assumes that the spoofer antenna is directional with $G_T^s = 10$ dB. The target receiver is 500 m away from the spoofer. The spoofing transmit power $P_T^s$ is tuned to be -65 dBm to make the arrival SAPR at the target receiver slightly higher than the presumed threshold. In this case, there is no overwhelming-spoofed area.

If the receiver stays outside the distance of $L_{TOV}$ from the spoofer, it is likely spoofed but not definitely. For example, if a receiver has been in a steady-state tracking of authentic signals at locations very far from the spoofer and gradually moves toward the spoofer, it will probably not be spoofed because the spoofing signals are still not overwhelming enough to force the receiver to loss lock. However, if the receiver loses locks on authentic signals due to signal blockage or other reasons, it will likely acquire/re-acquire the spoofing signals. If we denote the risky-spoofed range as $L_{TRI}$, it can be estimated by utilizing (2) as well with the substitution of the threshold $SAPR_{TRI}$, which is typically 3~5 dB.

In practice, there are two typical spoofing attack modes, respectively, the high-power-dominant attack and the covert approach-and-drag-off attack. A simulator-based spoofer or a meaconer normally apply the first spoofing mode and will dramatically affect the nearby receivers. Fig.1-(a) illustrates this type of spoofing scenario. On the other hand, the receiver-repeater is able to apply the second attack mode to the target receiver. Its transmitting power can be carefully tuned to make the arrival SAPR at the target receiver be slightly higher than the authentic signal. In this scenario, the spoofing power is normally too low to give rise to an overwhelming-spoofed area. However, it will still cause a risky-spoofed area, which is typically the extent between the spoofer and the target receiver, and the closer to the spoofer for the receivers, the more likely to be spoofed. Fig.1-(b) illustrates this type of spoofing scenario.

In this study, we try to detect the two types of spoofers shown in Fig.1 by utilizing the observable data of the crowdsourcing receivers. In the spoofing-free scenario, the pseudorange measurement $\rho_n^i$ from receiver $n$ for satellite $i$ is modeled as [45]:

$$\rho_n^i = |\mathbf{S}^i - \mathbf{p}_n| + c(\delta t_n - \delta t^i) + I_n^i + T_n^i + \varepsilon_n^i , \quad (3)$$

where $\mathbf{p}_n$ is receiver $n$'s position, $\delta t_n$ is receiver's local clock bias, $\mathbf{S}^i$ and $\delta t^i$ are satellite $i$'s position and clock bias, $c$ is the speed of light, $I_n^i$ and $T_n^i$ are the ionospheric and tropospheric delays, and $\varepsilon_n^i$ is the measurement noise of receiver $n$ for satellite $i$. $\varepsilon_n^i$ is assumed to be Gaussian with zero mean and a variance of $\sigma_\rho^2$.

Similarly, the pseudorange measurement $\rho_m^i$ from receiver $m$ for satellite $i$ can be obtained. With $\rho_n^i$ and $\rho_m^i$, we form a single differential pseudorange (SDP):

$$\Delta\rho_{nm}^i = \rho_n^i - \rho_m^i = |\mathbf{S}^i - \mathbf{p}_n| - |\mathbf{S}^i - \mathbf{p}_m| + c\Delta\delta t_{nm} + \Delta\varepsilon_{nm}^i , \quad (4)$$

where $\Delta\delta t_{nm} = \delta t_n - \delta t_m$ and $\Delta\varepsilon_{nm}^i = \varepsilon_n^i - \varepsilon_m^i$. The satellite clock bias, the ionospheric delay, and the tropospheric delay are cancelled in (4) since the distance between receiver $n$ and $m$ is typically within a few kilometers for the crowdsourcing scenario shown in Fig.1.

It is implicitly assumed in (4) that the measurement epochs for $\rho_n^i$ and $\rho_m^i$ of receiver $n$ and receiver $m$ are synchronized. However, some commercial receivers, e.g., certain types of smart phones, do not sample the measurements at synchronized time instants. In this case, (4) can be written as $\Delta\rho_{nm}^i(t_0, t_1) = |\mathbf{S}^i(t_0) - \mathbf{p}_n(t_0)| - |\mathbf{S}^i(t_1) - \mathbf{p}_m(t_1)| + c\Delta\delta t_{nm} + \Delta\varepsilon_{nm}^i$. As analyzed in [46], the asynchronized measurements from multiple receivers will produce a bias on the differential measurements which will affect the spoofing detection performance. Hence, necessary corrections, such as interpolation or pseudorange rate integration, need to be made to synchronize the measurements. Interested readers can refer to [46] for detailed information. In the following derivations, all the pseudorange measurements from different receivers are assumed to be properly synchronized.

According to [37,40], one more differential operation can be applied to (4) between different satellites to obtain a double differential pseudorange (DDP):

$$\nabla\Delta\rho_{nm}^{ij} = (|\mathbf{S}^i - \mathbf{p}_n| - |\mathbf{S}^i - \mathbf{p}_m|) - (|\mathbf{S}^j - \mathbf{p}_n| - |\mathbf{S}^j - \mathbf{p}_m|) + \nabla\Delta\varepsilon_{nm}^{ij} \ . \quad (5.a)$$

Since receiver $m$ is in a short baseline with respect to receiver $n$, (5.a) can be derived with Taylor expansions as:

$$\nabla\Delta\rho_{nm}^{ij} = (|\mathbf{S}^i - \mathbf{p}_n| - |\mathbf{S}^i - \mathbf{p}_n - \Delta\mathbf{p}_{nm}|) - (|\mathbf{S}^j - \mathbf{p}_n| - |\mathbf{S}^j - \mathbf{p}_n - \Delta\mathbf{p}_{nm}|) + \nabla\Delta\varepsilon_{nm}^{ij} \ \approx \ (\mathbf{e}_n^i - \mathbf{e}_n^j)^{\mathrm{T}}\Delta\mathbf{p}_{nm} + \nabla\Delta\varepsilon_{nm}^{ij} \ , \quad (5.b)$$

where $\mathbf{e}_n^*$ is the unit vector pointing from the receiver $n$ to satellite $*$, $\Delta\mathbf{p}_{nm} = \mathbf{p}_m - \mathbf{p}_n$, and the superscript T denotes transpose operation. By referring to [37], it has

$$\mathbf{e}_n^* = [\cos(\theta_n^*)\sin(\phi_n^*), \cos(\theta_n^*)\cos(\phi_n^*), \sin(\theta_n^*)]^{\mathrm{T}}, \quad (6)$$

where $\theta_n^*$ and $\phi_n^*$ are, respectively, the elevation and azimuth of satellite $*$ with respect to receiver $n$. Furthermore, we make $\mathbf{e}^i - \mathbf{e}^j = \left[e_x^{ij}, e_y^{ij}, e_z^{ij}\right]^{\mathrm{T}}$ to denote the elements of the differential geometry vector between satellite $i$ and $j$, and the receiver index $n$ is omitted for simplicity.

When receiver $n$ is spoofed, the counterfeit pseudorange $\rho_{n|\mathrm{f}}^i$ for the counterfeit satellite $i$ is expressed as

$$\rho_{n|\mathrm{f}}^i = |\mathbf{S}_\mathrm{f}^i - \mathbf{p}_{n|\mathrm{f}}| + c(\delta t_n - \delta t_\mathrm{f}^i) + I_{n|\mathrm{f}}^i + T_{n|\mathrm{f}}^i + L_{n|\mathrm{f}} + \varepsilon_n^i \ , \quad (7)$$

where the subscript f denotes the corresponding counterfeit terms manipulated by the spoofer, and $L_{n|\mathrm{f}}$ stands for the time-of-flight distance from the spoofer to the receiver $n$. If both receiver $n$ and $m$ are spoofed, it can readily be known that the DDP measurement will be $\nabla\Delta\rho_{nm}^{ij} \approx \nabla\Delta\varepsilon_{nm}^{ij}$, because of $\mathbf{p}_{n|\mathrm{f}} \approx \mathbf{p}_{m|\mathrm{f}} \approx \mathbf{p}_\mathrm{f}$ that is the counterfeit position. Then, a hypothesis detection method, such as GLRT, can be utilized to detect spoofing [37-38].

However, if receiver $n$ is spoofed and receiver $m$ is not, the DDP value will be

$$\nabla\Delta\rho_{n|\mathrm{f},m}^{ij} = (|\mathbf{S}_\mathrm{f}^i - \mathbf{p}_{n|\mathrm{f}}| - |\mathbf{S}^i - \mathbf{p}_m|) - (|\mathbf{S}_\mathrm{f}^j - \mathbf{p}_{n|\mathrm{f}}| - |\mathbf{S}^j - \mathbf{p}_m|) - \nabla\Delta\delta t_{n|\mathrm{f},m}^{ij} + \nabla\Delta I_{n|\mathrm{f},m}^{ij} + \nabla\Delta T_{n|\mathrm{f},m}^{ij} + \nabla\Delta\varepsilon_{nm}^{ij} \ , \quad (8.a)$$

where $\nabla\Delta\delta t_{n|\mathrm{f},m}^{ij}$, $\nabla\Delta I_{n|\mathrm{f},m}^{ij}$, and $\nabla\Delta T_{n|\mathrm{f},m}^{ij}$ are the corresponding double differential residual errors between the spoofing and the authentic signals. Usually, the spoofer tries to make its signal as close as to the authentic ones. Therefore, these residual errors are quite small and will be absorbed into the noise term $\nabla\Delta\varepsilon_{nm}^{ij}$. Besides, the terms $\delta t_n$ and $L_{n|\mathrm{f}}$ are cancelled by the double differential operation. In this case, the DDP value can be simplified as

$$\nabla\Delta\rho_{n|\mathrm{f},m}^{ij} \approx (\mathbf{e}_n^i - \mathbf{e}_n^j)^{\mathrm{T}}\Delta\mathbf{p}_{\mathrm{f}m} + \nabla\Delta\varepsilon_{nm}^{ij} \ , \quad (8.b)$$

where $\Delta\mathbf{p}_{\mathrm{f}m} = \mathbf{p}_m - \mathbf{p}_{n|\mathrm{f}} \approx \mathbf{p}_m - \mathbf{p}_\mathrm{f}$.

In the extreme case, suppose the residual errors cannot be neglected between spoofing and authentic signals, and we can derive (8.a) into the following expression:

$$\nabla\Delta\rho_{n|\mathrm{f},m}^{ij} = (|\mathbf{S}_\mathrm{f}^i - \mathbf{p}_\mathrm{f}| - |\mathbf{S}_\mathrm{f}^i - \mathbf{p}_\mathrm{f} - \Delta\mathbf{p}_{\mathrm{f}m} + \Delta\mathbf{S}^{\mathrm{f}i}|) - (|\mathbf{S}_\mathrm{f}^j - \mathbf{p}_\mathrm{f}| - |\mathbf{S}_\mathrm{f}^j - \mathbf{p}_\mathrm{f} - \Delta\mathbf{p}_{\mathrm{f}m} + \Delta\mathbf{S}^{\mathrm{f}j}|) - \nabla\Delta\delta t_{n|\mathrm{f},m}^{ij} + \nabla\Delta I_{n|\mathrm{f},m}^{ij} + \nabla\Delta T_{n|\mathrm{f},m}^{ij} + \nabla\Delta\varepsilon_{nm}^{ij}$$
$$\approx (\mathbf{e}_n^i - \mathbf{e}_n^j)^{\mathrm{T}}\Delta\mathbf{p}_{\mathrm{f}m} - \mathbf{e}_n^{i\mathrm{T}}\Delta\mathbf{S}^{\mathrm{f}i} + \mathbf{e}_n^{j\mathrm{T}}\Delta\mathbf{S}^{\mathrm{f}j} - \nabla\Delta\delta t_{n|\mathrm{f},m}^{ij} + \nabla\Delta I_{n|\mathrm{f},m}^{ij} + \nabla\Delta T_{n|\mathrm{f},m}^{ij} + \nabla\Delta\varepsilon_{nm}^{ij}$$
$$\approx (\mathbf{e}_n^i - \mathbf{e}_n^j)^{\mathrm{T}}\Delta\widetilde{\mathbf{p}}_{\mathrm{f}m} + \nabla\Delta\varepsilon_{nm}^{ij} \ , \quad (8.c)$$

where $\Delta\mathbf{S}^{\mathrm{f}*} = \mathbf{S}^* - \mathbf{S}_\mathrm{f}^*$, and $\Delta\widetilde{\mathbf{p}}_{\mathrm{f}m} = \mathbf{p}_m - \widetilde{\mathbf{p}}_\mathrm{f}$. It has $\widetilde{\mathbf{p}}_\mathrm{f} = \mathbf{p}_\mathrm{f} - (\mathbf{e}_n^{j\mathrm{T}}\Delta\mathbf{S}^{\mathrm{f}j} - \mathbf{e}_n^{i\mathrm{T}}\Delta\mathbf{S}^{\mathrm{f}i} - \nabla\Delta\delta t_{n|\mathrm{f},m}^{ij} + \nabla\Delta I_{n|\mathrm{f},m}^{ij} + \nabla\Delta T_{n|\mathrm{f},m}^{ij})/(\mathbf{e}_n^i - \mathbf{e}_n^j)^{\mathrm{T}}$. We can see that these unmatched model terms will result in an equivalent biased counterfeit position $\widetilde{\mathbf{p}}_\mathrm{f}$ instead of the original $\mathbf{p}_\mathrm{f}$. However, it will not affect the following derivation of the proposed algorithm.

From (8.a)~(8.c), it is known that the DP-GLRT method will fail to work in the case of one spoofed and the other not. In this study, we will try to use crowdsourcing receivers to solve this problem.

## IV. DERIVATION OF THE D²PS DISTRIBUTION

Without losing generality, we represent the dimension of the monitor area, i.e., the region of interest (ROI), by using the *East-North-Up* (ENU) local coordinate system. The monitor area in the $E$ direction spans from $a_1$ to $a_2$ ($a_1 < a_2$) and in the $N$ direction from $b_1$ to $b_2$ ($b_1 < b_2$). Since all the crowdsourcing receivers are considered on the ground, their coordinates in the $U$ direction are approximately the same. Next, we assume that in the monitor area there are $M$ crowdsourcing receivers, and each receiver has $J$ commonly observed satellite measurements. Based on the measurements of all the crowdsourcing receivers, the D²PS sample set is formed in the following way:

(1) Randomly choose a receiver as the reference and perform the pseudorange double differencing with all other receivers for a pair of satellite $i$ and $j$. It will produce $(M-1)$ double differential measurements.

(2) For the same pair of satellite, repeat the operation in step 1 by enumerating all receivers as the reference. This will result in $(M-1)M$ double differential measurements, which is referred to a subset of measurements, as $\mathbf{v}^{ij} = \left\{\nabla\Delta\rho_{m_n}^{ij}, \forall m_n \in (M-1)M\right\}$.

(3) Enumerate all satellite pairs, and repeat step 1 and 2. It will produce $C_J^2$ subsets, i.e. $\mathbf{v}^{ij}, \forall i_j \in C_J^2$, where $C_J^2$ denotes the number of combinations of choosing 2 from $J$, and each subset has $(M-1)M$ samples.

(4) Randomly permutate each subset $\mathbf{v}^{ij}$ and merge these subsets to form the D²PS sample set:

$$\text{D}^2\text{PS} = \left\{ \frac{1}{\sqrt{C_J^2}} \sum_{i_j=1}^{C_J^2} \text{randperm}(\boldsymbol{v}^{i_j}) \right\} =$$

$$\left\{ \frac{1}{\sqrt{C_J^2}} \sum_{i_j=1}^{C_J^2} \text{randperm}(\nabla\Delta\rho_{m_n}^{i_j}) \right\}, \forall m_n \in (M-1)M \ . \quad (9.a)$$

(5) If there are $K$ successive epochs of measurements, the final D$^2$PS samples can be further averaged among the sample sets at different epochs:

$$\text{D}^2\text{PS}_{(1:K)} = \frac{1}{K}\sum_{k=1}^{K} \text{D}^2\text{PS}_{(k)}[\text{idx}], \quad \forall \text{idx} \in (M-1)M \ . \quad (9.b)$$

We refer to the distribution of the samples from the D$^2$PS set formed above as the D$^2$PS distribution. According to the above steps, it can be seen that the authentic positions of the crowdsourcing receivers are not needed. The computation of the D$^2$PS sample set only relies on the reported pseudoranges of the receivers. Whether their measurements are fully or partially spoofed does not affect the work of this algorithm.

It is noticed that when deriving the D$^2$PS set the double differential measurements between a pair of receivers are computed twice for the same satellite pair subset $\boldsymbol{v}^{ij}$, but with opposite signs, say $\nabla\Delta\rho_{m_n}^{ij} = -\nabla\Delta\rho_{n_m}^{ij}$. Therefore, the elements in each subset $\boldsymbol{v}^{ij}$ is not independent. In order to mitigate this correlation, the random permutation in step (4) is introduced before the summation operation to make the elements with the same index in each subset mostly uncorrelated, especially when $M$ is a large number. Second, according to the central limit theorem, the element D$^2$PS[idx] will approximate to a normal distribution after the summation operation since $C_J^2$ is normally large, and those elements will approach to be uncorrelated with each other. Third, for different epochs, the number of common in-view satellites, $J$, may be different, which makes the D$^2$PS$_{(k)}$ derived from different sets of common in-view satellite of the receivers. However, the random permutation in (9.a) randomizes the D$^2$SP samples, making them less relevant to the specific satellite set. Thereby, the averaging operation in (9.b) will contribute to the D$^2$SP samples more approaching the theoretical statistical distribution. It will be further proved that the D$^2$PS set is a zero-mean normal distribution with a variance dependent on the specific spoofing scenario.

*A. Lemma-1:* In the spoofing-free scenario, the D$^2$PS distribution will asymptotically approach a zero-mean normal distribution as the number of receivers increases. Its probability function can be expressed as

$$p(\text{D}^2\text{PS})_{|a} \sim \mathcal{N}(0, \sigma_{\rho|a}^2), \quad (10.a)$$

where the subscript a stands for the spoofing-free scenario, and $\sigma_{\rho|a}^2$ is the variance, which can be estimated as

$$\sigma_{\rho|a}^2 = \frac{D_x^2}{6C_J^2} \sum_{i_j=1}^{C_J^2} \left(e_x^{ij}\right)^2 + \frac{D_y^2}{6C_J^2} \sum_{i_j=1}^{C_J^2} \left(e_y^{ij}\right)^2, \quad (10.b)$$

where $D_x = a_2 - a_1$ and $D_y = b_2 - b_1$ are the dimensions of the monitor area, and $e_x^{ij}$ and $e_y^{ij}$ are the x and y components of the differential geometry vector between satellite i and j.

The Proof of Lemma-1 is given in Appendix-A.

According to (10.b), the variance of the D$^2$PS samples in the spoofing-free scenario only depends on the dimension of the monitor area, the number of the visible satellites, and the differential geometry vectors between satellite pairs. These information are either a-prior known or available from public source.

*B. Lemma-2:* In the fully-spoofed scenario, the D$^2$PS distribution will follow a zero-mean normal function. Its probability function can be expressed as

$$p(\text{D}^2\text{PS})_{|f} \sim \mathcal{N}(0, \sigma_{\rho|f}^2), \quad (11.a)$$

where the subscript f stands for the fully-spoofed scenario, and $\sigma_{\rho|f}^2$ is the variance, which is approximately

$$\sigma_{\rho|f}^2 = 4\sigma_\rho^2, \quad (11.b)$$

where $\sigma_\rho^2$ is the variance of the pseudorange measurement error.

The Proof of Lemma-2 is given in Appendix-B.

*C. Lemma-3:* In the partially-spoofed scenario, the D$^2$PS distribution will asymptotically follow a zero-mean normal distribution. Its probability density function can be expressed as

$$p(\text{D}^2\text{PS})_{|af} \sim \mathcal{N}(0, \sigma_{\rho|af}^2), \quad (12.a)$$

where the subscript af denotes the partially-spoofed scenario, and $\sigma_{\rho|af}^2$ is the variance, which can be calculated as

$$\sigma_{\rho|af}^2 = (1-\alpha)^4 \sigma_{\nabla\Delta\rho_{mn}|a'}^2 + 4\alpha^4 \sigma_\rho^2 + 4\alpha^2(1-\alpha)^2 \sigma_{\nabla\Delta\rho_{lm}|s'}^2 + 2\alpha(1-\alpha)\left(\sigma_{\nabla\Delta\rho_{mn}|a'} + 2\sigma_\rho\right)\sigma_{\nabla\Delta\rho_{lm}|s'}, \quad (12.b)$$

where $\alpha$ is the percentage of the spoofed receivers among all crowdsourcing receivers, $\sigma_{\nabla\Delta\rho_{mn}|a'}^2$ is the variance of the DDP values between non-spoofed receivers, $4\sigma_\rho^2$ is the variance of the DDP values between spoofed receivers, $\sigma_{\nabla\Delta\rho_{lm}|s'}^2$ is the variance of the mutual differential measurements between spoofed and non-spoofed receivers. It can be proofed that

$$\sigma_{\nabla\Delta\rho_{mn}|a'}^2 = \frac{D_x^{a^2} \sum_{i_j=1}^{C_J^2}\left(e_x^{ij}\right)^2 + D_y^{a^2} \sum_{i_j=1}^{C_J^2}\left(e_y^{ij}\right)^2}{6C_J^2}, \quad (12.c)$$

where it is assumed that the non-spoofed receivers are uniformly spread in the horizontal area of $(a_1^a, a_2^a) \times (b_1^a, b_2^a)$, $D_x^a = a_2^a - a_1^a$ and $D_y^a = b_2^a - b_1^a$ are the length and width dimensions of the area, and

$$\sigma_{\nabla\Delta\rho_{lm}|s'}^2 = \frac{\sigma_{\Delta x_{lm}}^2 \sum_{i_j=1}^{C_J^2}\left(e_x^{ij}\right)^2 + \sigma_{\Delta y_{lm}}^2 \sum_{i_j=1}^{C_J^2}\left(e_y^{ij}\right)^2}{C_J^2}$$

$$\sigma_{\Delta x_{lm}}^2 = \frac{1}{3}[(x_f - a_1^a)^2 + (x_f - a_2^a)^2 + (x_f - a_1^a)(x_f - a_2^a)], \quad (12.d)$$

$$\sigma_{\Delta y_{lm}}^2 = \frac{1}{3}[(y_f - b_1^a)^2 + (y_f - b_2^a)^2 + (y_f - b_1^a)(y_f - b_2^a)]$$

where $(x_f, y_f)$ is the horizontal spoofing counterfeit position.

The Proof of Lemma-3 is given in Appendix-C.

Lemma-3 is a generalized extension for Lemma-1 and Lemma-2. If $\alpha = 0$, which corresponds to the spoofing-free scenario, $\sigma^2_{\rho|af}$ equals to $\sigma^2_{\nabla\Delta\rho_{mn}|a'}$, the variance of differential measurements of the non-spoofed receivers. Conversely, if $\alpha = 1$ that corresponds to the fully-spoofed scenario, $\sigma^2_{\rho|af}$ will be equal to $4\sigma^2_\rho$, the same as (11).

## V. Actual ROI Dimension Determination

In Section IV, the distribution of the D²PS sample set is derived based on the simplified assumption that the crowdsourcing receivers randomly scatter in the monitor region/ROI. However, in real applications, it may happen that the receivers locate in certain partial areas of the region. If we still use the whole region dimension, it will inevitably lead to the imprecise variance prediction for the computed D²PS sample. For this reason, a resizing technique of the monitor region is proposed to solve this problem.

- First, divide the entire ROI into a few equally smaller cell regions in a grid pattern. The dimension of the cell region is possibly a half, a quarter, or a fifth, etc., of the original region dimension. By default, the monitor ROI is a rectangular shape, and all these cell regions are rectangular as well.
- Second, map the receivers into these smaller regions according to their reported positions, and denote the number of receivers in a cell region as $M_{cr}$. Tag the cell region as an active cell only when the $M_{cr}$ is greater than a threshold.
- Third, group all the adjacent active cell regions as an enclosed region. Forming an enclosed region should follow three rules: (1) The enclosed region should always be a rectangular shape; (2) The enclosed region should contain adjacent cells; (3) An active cell region can be grouped into different enclosed regions as long as it meet the adjacent rule.
- Fourth, compute the D²PS sample set for each enclosed region, and take the dimension of each enclosed region as the corresponding size of the interest monitor area.

The minimal size of the cell region and the receiver density threshold are usually determined according to the specific application scenario. Fig.2 illustrates the resizing technique for a monitor region of 1 km x 1 km. Suppose for some reason the crowdsourcing receivers do not spread uniformly in this area. A spoofer lies in the north-west of the ROI and transmit the spoofing signal with a direction antenna toward the east. Suppose that the spoofer applies a surgical spoofing attack to a target, which means that the transmit power is moderate. Therefore, only some receivers lying close to the spoofed are spoofed and most of the receivers that stay far are not affected. The counterfeit position conceived by the spoofer is at the south-west of the ROI. Due to the hardware diversity, the reported positions of these spoofed receivers will scatter around the conceived counterfeit position (illustrated by the purple triangles).

If the dimension of the entire ROI is directly utilized to estimate the variance of the D²PS samples, there will inevitably a large prediction error. Whereas, by utilizing the resizing technique, the ROI can be divided into 25 cell regions, each of which has a dimension of 200 m. After grouping adjacent active cell regions, two enclosed regions are obtained. The two D²PS sample subsets for the enclosed regions can be computed, respectively. Fig.3 shows the histograms of the D²PS subset samples and the corresponding theoretical distributions for these two enclosed regions. It is found that for enclosed region 1 the histogram of the D²PS subset samples (shown by blue bars) matches the theoretical distribution (shown by the dashed red curve), that is predicted by assuming the case of the spoofing-free situation, although the spoofing indeed happens in this region. This is because the measurements of the spoofed receivers are not involved in the computation of the D²PS samples of enclosed region 1 since they are mapped into the enclosed region 2 according to their reported positions. On the contrary, for the enclose region 2, it is clearly seen that the histogram of the samples deviates from the theoretical spoofing-free distribution. This deviation can be used for spoofing detection.

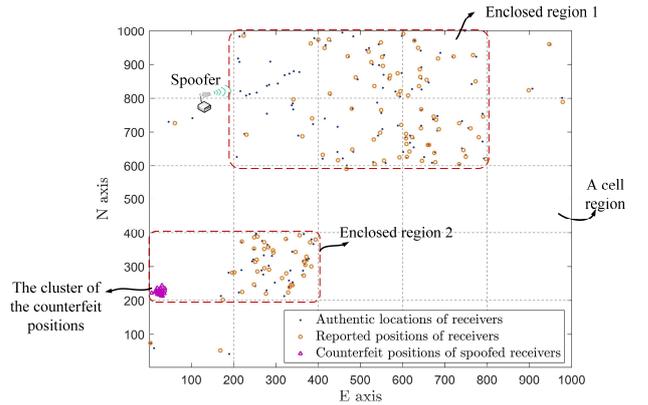

Fig.2 An illustration of a spoofing scenario in which the crowdsourcing receivers are not spreading uniformly in the ROI and the resizing technique is applied to obtain the enclosed regions.

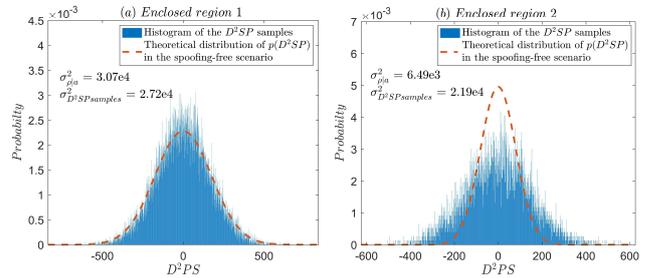

Fig.3 The histogram of the D²SP sample subsets as well as the theoretical distributions of the spoofing-free scenario for the two enclosed regions shown in Fig.2.

## VI. D²PS Sample Variance as A Spoofing Indicator

In Section IV, it has been demonstrated that the variance of the D²PS sample set is distinct for different spoofing scenarios. Therefore, it can be used as a statistic metric for spoofing detection. Let us denote the D²PS sample set as $\text{D}^2\text{PS} = \{d_1, d_2, ..., d_N\}$, where $N$ is the number of the samples in the set. The variance of the D²PS distribution can be estimated by

$$\sigma^2_{\text{D}^2\text{PS}} = \frac{1}{N-1}\sum_{n=1}^{n=N} d_n^2 . \quad (13)$$

We define the spoofing-free case as $H_0$ hypothesis, the fully-spoofed scenario as $H_1$ hypothesis, and the partially-spoofed scenario as $H_2$ hypothesis. According to the derivations of Lemma-1, -2, and -3, it can be inferred that the variance in $H_1$ is usually smaller than that in $H_0$ because the dimensions $D_x$ and $D_y$ of the ROI is typically much larger than the magnitude of the pseudorange errors. The variance of $H_2$ is more complicated. If the percentage of the spoofed receivers $\alpha$ is close to 100%, the variance of $H_2$ will approach that of $H_1$. Conversely, if $\alpha$ close to 0, the variance of $H_2$ will approach that of $H_0$. The variance of $H_2$ is also affected by the counterfeit spoofed receiver position $\boldsymbol{p}_\text{f}$. If $\boldsymbol{p}_\text{f}$ is farther away from the ROI, the variance of $H_2$ will be larger than that of $H_0$. On the contrary, if $\boldsymbol{p}_\text{f}$ is within or near the monitoring area, the variance of $H_2$ is close to that of $H_0$, which will make it harder to distinguish $H_0$ and $H_2$ hypothesizes.

Based on the above analysis, a tri-level hypothesis decision rule is developed as follows:

$$\begin{aligned} H_0: \gamma_1 \leq \sigma^2_{\text{D}^2\text{PS}} \leq \gamma_2 \\ H_1: \sigma^2_{\text{D}^2\text{PS}} < \gamma_1 \\ H_2: \sigma^2_{\text{D}^2\text{PS}} > \gamma_2 \end{aligned} \quad (14)$$

For the $H_0$ hypothesis, it is known that $M \sigma^2_{\text{D}^2\text{PS}}/\sigma^2_{\rho|a} \sim \chi^2_M$ follows a chi-squared distribution with $M$ degrees of freedom [50,51]. It needs to pointed out that the freedom degree for the chi-squared distribution of (14) is $M$ rather than $N-1$ because the samples in the D²PS set is formed by $M$ independent receivers.

Next, define the one-sided false alarm probability for deciding $H_1$ be $\epsilon/2$ given that the $H_0$ hypothesis is true. Then it has

$$F_{\chi^2_M}\left(\frac{M}{\sigma^2_{\rho|a}}\gamma_1\right) = \epsilon/2 , \quad (15)$$

where $F_{\chi^2_M}(x)$ is the cumulative function of the chi-squared distribution $\chi^2_M$. Hence, the threshold $\gamma_1$ can be obtained by

$$\gamma_1 = \frac{\sigma^2_{\rho|a}}{M} F^{-1}_{\chi^2_M}(\epsilon/2) . \quad (16)$$

Here, $F^{-1}_{\chi^2_M}(y)$ is the inverse of the chi-squared cumulative function.

Correspondingly, set the one-sided false alarm probability for deciding $H_2$ to $\epsilon/2$ given $H_0$ is true, then it has

$$F_{\chi^2_M}\left(\frac{M}{\sigma^2_{\rho|a}}\gamma_2\right) = 1 - \epsilon/2 , \quad (17)$$

and the threshold $\gamma_2$ can be computed by

$$\gamma_2 = \frac{\sigma^2_{\rho|a}}{M} F^{-1}_{\chi^2_M}(1-\epsilon/2) . \quad (18)$$

The detailed expression for the functions $F_{\chi^2_M}(x)$ and $F^{-1}_{\chi^2_M}(x)$ can be referred to [50,51]. It can be readily seen that the overall false alarm probability is $\epsilon$ for deciding either $H_1$ or $H_2$ when $H_0$ is true.

Once the thresholds $\gamma_1$ and $\gamma_2$ are determined, the final detection probabilities $P_{d|H_1}$ and $P_{d|H_2}$ can be derived, respectively, as

$$\begin{aligned} P_{d|H_1} = F_{\chi^2_M}\left(\frac{M}{\sigma^2_{\rho|f}}\gamma_1\right) \\ P_{d|H_2} = 1 - F_{\chi^2_M}\left(\frac{M}{\sigma^2_{\rho|af}}\gamma_2\right) \end{aligned}. \quad (19)$$

## VII. Performance Evaluation

In this section, the performance of the spoofing detection by the D²PS sample variance is investigated under various scenarios, in comparison with the DP-GLRT spoofing detection method (or GLRT for short) that is elaborated in [38-40]. GPS ephemeris collected on June 21$^{th}$ 2021 in Shanghai is used in the simulation tests. There are 12 GPS satellites above 10$^o$ elevation mask during this time. The pseudorange measurements of all the receivers under different spoofing scenarios are simulated, in that an estimator-repeater spoofer is assumed. Hence, the satellite positions, the ionospheric and the tropospheric delays in the spoofing signals are almost the same as the authentic ones. A 500-nanosecond processing induced by the spoofer hardware delay is also considered in the simulation model.

### A. Detection performance for fully-spoofed scenarios

#### a.1) Fully-spoofed scenario in an open sky environment

In the first test, the receiver operating characteristics (ROC) performances of the D²PS and DP-GLRT methods in the fully-spoofed scenario are evaluated and compared in Fig.4. There suppose $M$ receivers randomly scattering in a $D \times D$ ROI, where $D$ is either 50 m or 100 m, and $M$ is either 10 or 20. The number $J$ of the common in-view satellites among the $M$ receivers is 12. For D²PS, its decision thresholds, $\gamma_1$ and $\gamma_2$, are determined according to the formulas given in Section-VI. For DP-GLRT, its decision threshold is determined by referring to [39,40]. For each test case, 1000 Monte Carlo simulation runs are performed to get the corresponding $H_1$ detection probability $P_{d_{H_1}}$ under a predefined false alarm (FA) rate for the $H_0$ scenario. The standard deviation, $\sigma_\rho$, of the pseudorange measurement errors is assumed to be 5 m, corresponding to an open-sky environment with a normal level of user equivalent range error (UERE) [45].

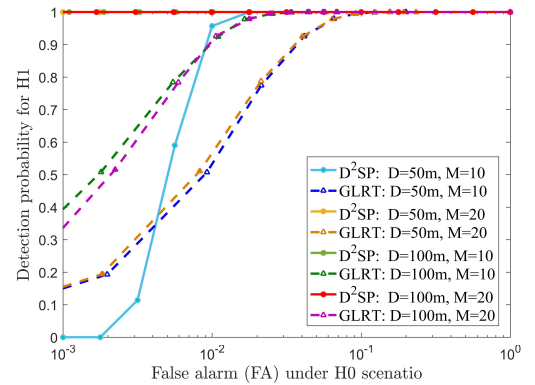

Fig.4 ROC performance comparison between D²PS (solid lines marked with dots) and DP-GLRT (dashed lines marked with triangles) with different ROI dimensions and receiver numbers.

It is seen that when the ROI dimension is small and the receiver number is few (*e.g.* the case of *D*=50 m and *M*=10), the D²PS method obtains a poor ROC performance (the blue solid curve with dots). Especially, it cannot detect spoofing when the FA is low to 0.001. However, its ROC curve is significantly improved when the ROI dimension *D* is enlarged to 100 m or the number *M* of receivers is increased to 20. This result coincides with the prediction of Lemma-1 that the distribution of D²PS in the spoofing-free scenario will be more distinct from that in the fully-spoofed scenario when the ROI area is larger. Moreover, when the receiver number is greater, the computed D²PS variance is more close to the theoretically predicted value, thus resulting to a more accurate detection decision.

As to DP-GLRT, its ROC performance improves as the dimension of the ROI grows. This is reasonable because the DDP value as given by (5.b) is larger when the pair of receivers are more distantly separated, thus improving the detection probability and lowering the FA. It is also found that the ROC curve is independent on the number of receivers. This is because the DP-GLRT method performs the hypothesis detection once for a pair of receivers and conclude the final decision by counting the ratio of truth hypothesis. Therefore, the detection probability closely depends on the relative distance of the two receivers. When the ROI dimension is fixed, the increase of the receiver number will not change the ratio of the relative ranges of the receivers, thus the same for the ratio of the truth decisions.

Generally speaking, in a realistic spoofing scenario, the dimension of the ROI and the number of the receivers are usually far greater than $D = 50$ and *M*=10. It is suggested that the D²PS method better work with the ROI size of $D \geq 100$ m and the receiver number $M \geq 20$ to achieve a satisfactory performance.

*a.2) Fully-spoofed scenario in an urban environment*

In the second test, we simulate a fully-spoofed scenario in an urban environment, where there are large multipath-induced errors in the pseudorange measurements. The multipath error model suggested by the Minimum Operational Performance Standards (MOPS) for GPS/SBAS Airborne Equipment [52] is adopted in the simulation. This model assumes a Gauss-Markov process with a variance of $\sigma_{mp}^2$ and a correlation time $\beta_{mp}^{-1}$ to characterize the random walk nature of the error. The variance $\sigma_{mp}^2$ relates to the elevation $\theta$ of the incident signal:

$$\sigma_{mp}^2 = \delta_\sigma^2\big(0.13 + 0.53\exp(-\theta/10\deg)\big), \quad (20)$$

where $\delta_\sigma$ is the inflation factor to reflect the severity of the multipath error in an urban environment. For an authentic signal, the parameter $\theta_a$ corresponds to its real satellite elevation. For a spoofing signal, $\theta_s$ is assumed to be 5 degrees. The correlation time $\beta_{mp}$ is set to be 25 seconds according to [52].

Apart from the random walk nature of the multipath error, the correlation of the errors between different receivers needs to be considered as well because they are in the same multipath environment. We assume that the correlation coefficient of the multipath errors for two receivers depends on their relative distance. That is

$$c_{nm} = \exp(-|d_{nm}|/\tau_d), \quad (21)$$

where $d_{nm}$ denotes the distance between receiver *n* and *m*, and $\tau_d$ is the correlation decay parameter that is assume to be 25 m in the simulation. Then, a correlation matrix $\boldsymbol{C}$ between all pairs of receivers can be obtained as

$$\boldsymbol{C} = \boldsymbol{R}^{\mathrm{T}}\boldsymbol{R} = [c_{nm}]_{1 \leq n \leq N, 1 \leq m \leq N}, \quad (22)$$

where the matrix $\boldsymbol{C}$ is symmetric and positive semi-definite by its definition, and $\boldsymbol{R}$ is the upper triangular square root matrix of $\boldsymbol{C}$. Last, the multipath errors of all the receivers are multiplied by $\boldsymbol{R}$ to construct the mutual correlation-ship among the errors.

Fig.5 shows the ROC performance comparison between D²PS and DP-GLRT with different multipath error inflation factors. $\delta_\sigma$ is set to 5, 10, 15, and 20, respectively. The larger the factor, the severe the multipath. In addition to the multipath error, the UERE error with the standard deviation $\sigma_\rho$ = 5 m still exist in the measurements. The ROI dimension *D* is 100 m and the crowdsourcing receiver number *M* is 20.

One can observe from Fig.5 that the performance of D²PS

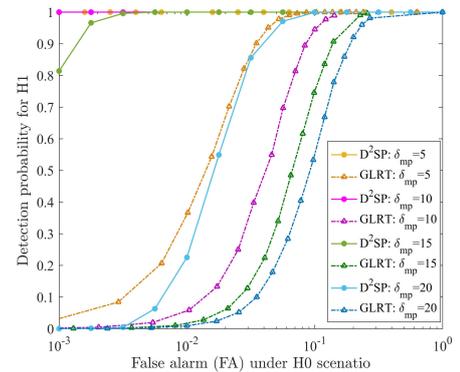

Fig.5 ROC performance comparison between D²PS (solid lines marked with dots) and DP-GLRT (dashed lines marked with triangles) under different urban multipath scenarios. The ROI dimension *D* is 100 m and the crowdsourcing receiver number *M* is 20.

outperforms that of DP-GLRT. When the multipath inflation factor $\delta_\sigma$ is moderate, such as 5 and 10. The D$^2$PS is almost immune to the effects of multipath. Its $P_{d_{H_1}}$ reaches almost 100% even when the FA is kept as low as 0.001. Until $\delta_\sigma$ is increased up to 20, the ROC curve of D$^2$PS is noticeably degraded. As to DP-GLRT, it is readily seen that it is more vulnerable to multipath errors.

*B. Detection performance for partially-spoofed scenarios*

*b.1) Detection performance with partial receivers spoofed*

In the third test, a scenario with partial receivers spoofed is simulated. The total receiver number is set to 100 and the size of the ROI is $D$=1000 m. The percentage $\alpha$ of the spoofed receivers is varied from 0.05 to 0.95. Three cases, respectively, $|\boldsymbol{p}_f|/D$=0.8, $|\boldsymbol{p}_f|/D$=1, and $|\boldsymbol{p}_f|/D$=1.5, are simulated, where the value $|\boldsymbol{p}_f|$ is the distance of the counterfeit spoofing position from the local coordinate origin. The height of the counterfeit position $\boldsymbol{p}_f$ is roughly the same as the ROI, whereas the azimuth of the $\boldsymbol{p}_f$ is randomly distributed with respect to the origin. The environment is assumed to be open-sky with the pseudorange measurement standard deviation $\sigma_\rho$ equal to 5 m.

According to [38-40], the original DP-GLRT algorithm was not designed to work with the partially-spoofed scenario. In order to make it comparable with D$^2$PS, we apply the GLRT method once for a pair of satellites of two receivers, and repeat the method for all satellites pairs and receiver pairs. Finally, there are a total of $C_J^2 C_M^2$ detection results. If at least 90% of individual detection results claim $H_1$ hypothesis, a global $H_1$ hypothesis is decided true. If at least 90% of individual detection results claim $H_0$ hypothesis, a global $H_0$ hypothesis will be the final result. Otherwise, the $H_2$ hypothesis is claimed.

It can be seen from Fig.6 that the detection performance of the D$^2$PS method for the $H_2$ scenario is greatly affected by the spoofing counterfeit position $\boldsymbol{p}_f$. As the counterfeit position still lies in the ROI, such as $|\boldsymbol{p}_f|/D$ =0.8, the detection probability can only reach maximally over 50% when the spoofed ratio $\alpha$ is between 0.4~0.5. However, when the counterfeit position is noticeably far from the ROI, like $|\boldsymbol{p}_f|/D$=1.5, the detection probability is significantly improved to almost 100% when the spoofed ratio $\alpha$ is below 0.9. As more receivers are spoofed, the variance of the D$^2$PS samples will go down toward the level of a fully-spoofed scenario. This behavior can be explained by Lemma-3 since both the counterfeit position and the spoofed ratio are involved in the variance computation.

The detection performance of the DP-GLRT method for the $H_2$ scenario does not depend on the counterfeit spoofing position $\boldsymbol{p}_f$, but on the percentage of spoofed receivers $\alpha$. When $\alpha$ <0.3, DP-GLRT will always claim the $H_2$ scenario as a $H_0$ hypothesis. This phenomenon relates to the rule that DP-

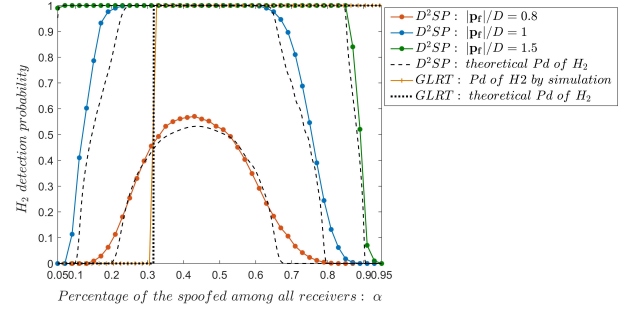

Fig.6 Detection probability comparisons for the partially-spoofed scenario ($H_2$) with different $|\boldsymbol{p}_f|/D$ ratios. The ratio $\alpha$ of the spoofed ones among all receivers varies from 0.05 to 0.95, where the receiver number is set as $M$=100. The ROI size is set to $D$=1000m. For each test case, 1000 Monte Carlo simulation runs are performed to get the detection probability for each method.

GLRT makes a decision depending on the percentage of the individual decisions on $H_1$ hypothesis among all results. This percentage can be predicted by

$$T_{H_1,\text{GLRT}} = \frac{C_J^2 C_{\alpha M}^2}{C_J^2 C_M^2} = \frac{\alpha\left(\alpha - \frac{1}{M}\right)}{1 - \frac{1}{M}} \ . \quad (23)$$

When $M \gg 1$, it has $T_{H_1,\text{GLRT}} \approx \alpha^2$. Thus, if $T_{H_1,\text{GLRT}} >0.1$, it requires $\alpha >0.316$. If $T_{H_1,\text{GLRT}} >0.9$, it requires $\alpha >0.949$. As shown in Fig.6, the detection performance of GLRT by Monte Carlo simulations matches well the predicted performance curve given by (23).

*b.2) Detection performance with partial satellites spoofed*

In the fourth test, another type of partially-spoofed scenario is simulated, in which a fraction of the in-view satellites is spoofed by the spoofer. In this test, the size of the ROI area is set to 100 m and 500 m, respectively. The number of the receivers is set to 25 and 50, respectively. The number of spoofed satellites varies from 4 to 12. The FA is fixed as 0.001 to determine the corresponding thresholds. The corresponding spoofing detection probabilities of D$^2$PS and DP-GLRT are displayed in Fig.7.

It is seen that the $P_d$ of D$^2$PS grows as the number of spoofed satellites increases. It is interesting to note that the performance of D$^2$PS depends on the number of receivers, $M$, instead of the size of area, $D$. This is because the variance of D$^2$PS in the partial-satellite-spoofed scenario is a little different from that of partial-receiver-spoofed scenario. According to Lemma-3, its variance can be derived as

$$\sigma^2_{\rho|\text{af}} = \left(1 - \frac{C_{J_s}^2}{C_J^2}\right)^2 \sigma^2_{\nabla\Delta\rho_{mn}|\text{a}'} + 4\left(\frac{C_{J_s}^2}{C_J^2}\right)^2 \sigma^2_\rho \ . \quad (24)$$

The mutual standard deviation term $\sigma_{\nabla\Delta\rho_{lm}|s'}$ is eliminated by common in-view authentic and spoofing satellites between different receivers. Although $\sigma^2_{\rho|\text{af}}$ in (24) does not depends on the receiver number $M$ but on the area size $D$, the threshold $\gamma_1$ is determined by normalizing $\sigma^2_{\rho|\text{af}}$ with $M$, as shown in (15).

The change of $D$ will make the threshold change accordingly due to the fixed FA. However, a larger $M$ will make the sample variance more precisely approach the theoretic value. Therefore, the detection accuracy is improved. From Fig.7, it is seen the D²PS can achieve a satisfactory performance when the ratio of spoofed satellites is greater than 75% with the number of receivers is large.

On the contrary, Fig.7 shows that the detection performance of DP-GLRT relates to the area size $D$ rather than the receiver number $M$. The reason for this is the same as that explained for the test result in Fig.4: As for the same ROI dimension, the increase of the receiver number will not change the ratios of the relative range magnitudes among the receivers. Therefore, the ratio of the truth decisions that determine the global detection will not change as well. Nevertheless, it can be readily seen that the detection performance of D²PS is better than that of DP-GLRT under this partially-spoofed scenario.

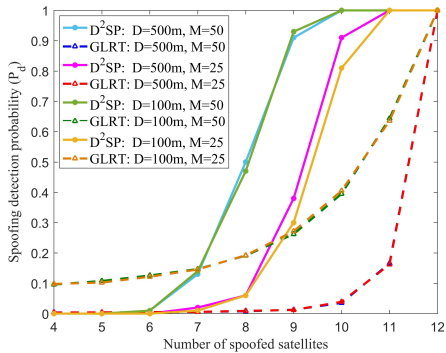

Fig.7 Detection probability for the scenario of partial satellites spoofed. The number of spoofed satellites varies from 4 to 12, while the total in-view satellites are 12. For each test case, 1000 Monte Carlo simulation runs are performed to get the detection probability for each method.

### C. Computational efficiency comparisons

Another advantage of the D²SP method over the DP-GLRT method is its computational efficiency. Both methods rely on the double differential pseudorange measurements of the crowdsourcing receivers. Suppose that there are $M$ receivers, $J$ commonly observed satellites, and $K$ successive observation epochs. We need $KJC_J^2C_M^2$ subtractions to obtain all the double differential measurements for both methods. For D²SP, there will need $KM(M-1)C_J^2$ additions to obtain the D²SP sample sets, and additional $M(M-1)$ multiplications and additions to get the final variance of the D²SP samples. In total, there are approximately $\frac{1}{2}KM^2J^2\left(1+\frac{J}{2}\right)$ equivalent additions and $M^2$ multiplications.

The DP-GLRT requires $C_J^2C_M^2$ individual GLRT processes to obtain the final decision. For a single GLRT, there will need approximately $4K(K+2)$ additions, $4K(K+2.5)$ multiplications, and two divisions. There will be approximately $K^2M^2J^2$ equivalent additions, $K^2M^2J^2$ multiplications, and $\frac{1}{2}M^2J^2$ divisions of computations in total. Obviously, the computation efficiency of the D²SP method is superior to the DP-GLRT method, especially when $K$, $M$, and $J$ are large.

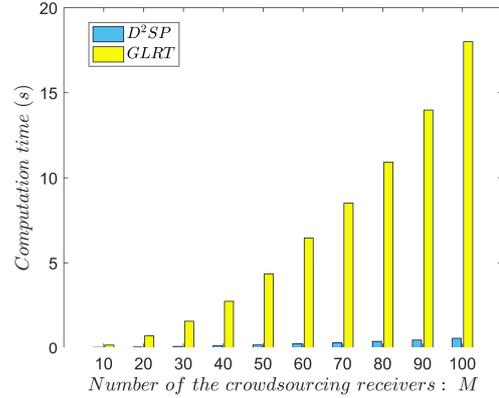

Fig.8 The computational time comparison between D²SP and GLRT. The dimension of the ROI is 1000 m x 1000 m, $J$=12, and $K$=5. A 1000-times Monte-Carlo simulation is used for each method to obtain the corresponding average time.

Fig.8 shows the comparison of the computation time of D²SP and DP-GLRT. Still, it is assumed a ROI with a dimension of 1000 m x 1000 m, the commonly observed satellites number is $J$=12, and the successive observation epochs is $K$=5. We change the crowdsourcing receiver number $M$ from 10 to 100, and evaluate the corresponding average computation time of each method by 1000 Monte-Carlo simulation runs. The Matlab-based script codes are run on a laptop with an Intel i7 CPU. It is seen that the computation time consumed by DP-GLRT grows dramatically with the increase of $M$, whereas the time used by D²SP only grow very moderately.

### VIII. CONCLUSION

In this paper, a new spoofing detection algorithm based on crowdsourcing receivers' double differential pseudorange measurements spatial distribution is proposed. The process of constructing the D²SP random set is presented and its distributions under different spoofing scenarios are derived. Based on the variance of the D²SP, a tri-level hypothesis detection test is designed to classify the scenarios of spoofing-free, fully-spoofed, and partially-spoofed of the ROI. It does not require the prior knowledge of the truth positions or relative distances of the receivers. Compared with the conventional DP-GLRT, the proposed algorithm shows better tolerance for the multipath errors and computational efficiency. It also shows better performance in the partially-spoofed scenario than DP-GLRT. Besides, the D²SP algorithm can easily adapt itself for different sizes of ROI and numbers of crowdsourcing receivers. The method can be applied to different GNSS signals and different frequency bands that can help find in which signals or bands the spoofing occurs. It is expected that the carrier phase or the Doppler data will contribute to the detection performance if they can be properly used. These works will be focused in the future research.

APPENDIX-A: PROOF OF LEMMA-1

Denote the position of receiver $m$, in the local ENU coordinate system, as $\boldsymbol{p}_m = [x_m, y_m, z_m]^\mathrm{T}, \forall m \in M$. Since the receiver is assumed to be placed randomly in the monitoring area, $x_m \sim \mathcal{U}(a_1, a_2)$ and $y_m \sim \mathcal{U}(b_1, b_2)$, where $\mathcal{U}(a_1, a_2)$ stands for a uniform distribution in $[a_1, a_2]$. If receiver $n$ is taken as the reference, which is also randomly placed in the area, the position difference between receiver $m$ and $n$ is $\Delta \boldsymbol{p}_{nm} \approx [\Delta x_{nm}, \Delta y_{nm}, 0]^\mathrm{T}$. It can be readily shown that $\Delta x_{nm} \sim \mathcal{U}(a_1 - x_n, a_2 - x_n)$ and $\Delta y_{nm} \sim \mathcal{U}(b_1 - y_n, b_2 - y_n)$. When the reference receiver is enumerated from 1 to $M$, a family of uniform distributions will be obtained:

$$\begin{array}{ll} \Delta x_{nm} \sim \mathcal{U}(a_1 - x_n, a_2 - x_n)|_{m \in M, n \in M} & \forall x_n \sim \mathcal{U}(a_1, a_2) \\ \Delta y_{nm} \sim \mathcal{U}(b_1 - y_n, b_2 - y_n)|_{m \in M, n \in M} & \forall y_n \sim \mathcal{U}(b, b_2) \end{array}. \quad (A.1)$$

Let us first examine the distribution of $\Delta x_{nm}$. Define the assembly of the uniform distribution families $\Delta x_{nm}$ as $\Delta X = \{\Delta x_{nm}\}, \forall m \in M, n \in M$. It is known that each uniform distribution in the family has the same support range of $a_2 - a_1$ but different midpoint values. The midpoint uniformly spreads from $-(a_2 - a_1)/2$ to $(a_2 - a_1)/2$ because of $x_n \sim \mathcal{U}(a_1, a_2)$. Thus, the samples in $\Delta X$ will follow a triangle distribution as [47]

$$\Delta X \sim Tr(tr_a = -D_x, tr_b = D_x, tr_c = 0) = \begin{cases} \frac{\Delta x + D_x}{D_x^2} & -D_x \leq \Delta x \leq 0 \\ \frac{D_x - \Delta x}{D_x^2} & 0 \leq \Delta x \leq D_x \\ 0 & \Delta x < -D_x \text{ or } > D_x \end{cases}, (A.2)$$

where $tr_a$, $tr_b$, and $tr_c$ denote the minimum, maximum, and peak of a triangle distribution, respectively. Similarly, the assembly of random samples, $\Delta Y = \{\Delta y_{nm}\}, \forall m \in M, n \in M$, will also follow a triangle distribution as $\Delta Y \sim Tr(tr_a = -D_y, tr_b = D_y, tr_c = 0)$.

In (9.b), the first term, $\left(\boldsymbol{e}_n^i - \boldsymbol{e}_n^j\right)^\mathrm{T} \Delta \boldsymbol{p}_{nm}$, can be written as $e_x^{ij} \Delta x_{nm} + e_y^{ij} \Delta y_{nm}$, which is the sum of two scaled triangle-distributed random values. Deriving its distribution needs the inverse Fourier transform of the product of the characteristic functions of $\Delta x_{nm}$ and $\Delta y_{nm}$. By referring to [48,49], the closed form of the probability function of the term $e_x^{ij} \Delta x_{nm} + e_y^{ij} \Delta y_{nm}$ can be found to be a symmetric piecewise function:

$$p\left(h \triangleq e_x^{ij} \Delta x_{nm} + e_y^{ij} \Delta y_{nm}\right) = \begin{cases} k_{tr} p_1(|h|) & 0 \leq |h| < D_{e_x}^{ij} - D_{e_y}^{ij} \\ k_{tr} p_2(|h|) & D_{e_x}^{ij} - D_{e_y}^{ij} \leq |h| < D_{e_y}^{ij} \\ k_{tr} p_3(|h|) & D_{e_y}^{ij} \leq |h| < D_{e_x}^{ij} \\ k_{tr} p_4(|h|) & D_{e_x}^{ij} \leq |h| < D_{e_x}^{ij} + D_{e_y}^{ij} \end{cases}, \quad (A.3)$$

where $k_{tr} = 1/\left(D_{e_x}^{ij} D_{e_y}^{ij}\right)^2$, $D_{e_x}^{ij} = e_x^{ij} D_x$, $D_{e_y}^{ij} = e_y^{ij} D_y$, and $p_1(|h|) \sim p_4(|h|)$ are given in (A.4).

When deriving (A.3) and (A.4), it is assumed that $D_{e_x}^{ij} > D_{e_y}^{ij}$ and $D_{e_y}^{ij} > D_{e_x}^{ij} - D_{e_y}^{ij}$. This assumption will not alter the following analysis result although the detailed expression in (A.3) and (A.4) will be different if the assumption changes. Furthermore, it can be known that the mean of $p(h)$ is 0, and its variance is

$$\sigma_h^{ij^2} = \frac{\left(D_{e_x}^{ij}\right)^2 + \left(D_{e_y}^{ij}\right)^2}{6}. \quad (A.5)$$

In a fact, $\left(D_{e_x}^{ij}\right)^2/6$ and $\left(D_{e_y}^{ij}\right)^2/6$ are, respectively, the variances of the scaled triangle-distributed random value $e_x^{ij} \Delta x_{nm}$ and $e_y^{ij} \Delta y_{nm}$. Since $\Delta x_{nm}$ and $\Delta y_{nm}$ are uncorrelated, the variance of the sum of the two scaled triangle-distributed random values is just the sum of their variances.

In real applications, the magnitude of the error term in (9) is much smaller than the dimension of the ROI, i.e., $\left|\nabla \Delta \varepsilon_{nm}^{ij}\right| < D_x + D_y$. Therefore, it has $\nabla \Delta \rho_{nm}^{ij} \approx \left(\boldsymbol{e}_n^i - \boldsymbol{e}_n^j\right)^\mathrm{T} \Delta \boldsymbol{p}_{nm} \approx e_x^{ij} \Delta x_{nm} + e_y^{ij} \Delta y_{nm}$. Hence, the distribution of the subset $\boldsymbol{v}^{ij}$ is just $p(h)$.

When all the differential satellite pairs are enumerated and all the measurement subsets are merged, each sample in the final D²PS set is the sum of $C_J^2$ random values whose probability distributions are the same as $p(h)$ but with different variances. Because these measurement subsets are independent realizations for their corresponding random process, according to the central limit theorem, the final probability distribution of the D²PS set will asymptotically approach a normal function, as given in (12.a), and its variance will be the sum of the variances of the subsets:

$$\begin{array}{l} p_1(|h|) = \frac{|h|^3}{3} - 3|h|^2 D_{e_y}^{ij} + \frac{\left(3D_{e_x}^{ij} - D_{e_y}^{ij}\right)}{3}\left(D_{e_y}^{ij}\right)^2 \\ p_2(|h|) = \frac{|h|^3}{2} - \frac{\left(D_{e_x}^{ij} + D_{e_y}^{ij}\right)}{2}|h|^2 + \frac{\left(D_{e_x}^{ij} - D_{e_y}^{ij}\right)^2}{2}|h| + \frac{\left(3D_{e_y}^{ij} - D_{e_x}^{ij}\right)}{6}\left(D_{e_x}^{ij}\right)^2 + \frac{\left(3D_{e_x}^{ij} - D_{e_y}^{ij}\right)}{6}\left(D_{e_y}^{ij}\right)^2 \\ p_3(|h|) = \frac{|h|^3}{6} + \frac{\left(D_{e_y}^{ij} - D_{e_x}^{ij}\right)}{2}|h|^2 + \frac{\left(\left(D_{e_x}^{ij}\right)^2 - \left(D_{e_y}^{ij}\right)^2 - 2D_{e_x}^{ij} D_{e_y}^{ij}\right)}{2}|h| + \frac{\left(3D_{e_y}^{ij} - D_{e_x}^{ij}\right)}{6}\left(D_{e_x}^{ij}\right)^2 + \frac{\left(3D_{e_x}^{ij} + D_{e_y}^{ij}\right)}{6}\left(D_{e_y}^{ij}\right)^2 \\ p_4(|h|) = -\frac{|h|^3}{6} + \frac{\left(D_{e_x}^{ij} + D_{e_y}^{ij}\right)}{2}|h|^2 - \frac{\left(D_{e_x}^{ij} + D_{e_y}^{ij}\right)^2}{2}|h| + \frac{\left(D_{e_x}^{ij} + D_{e_y}^{ij}\right)^3}{6} \end{array} \quad (A.4)$$

$$\sigma_{\rho|a}^2 = \frac{1}{C_j^2}\sum_{i_j=1}^{C_j^2}\sigma_h^{i_j^2} = \frac{\left(D_{e_x}^{i_j}\right)^2+\left(D_{e_y}^{i_j}\right)^2}{6C_j^2}$$
$$= \frac{D_x^2}{6C_j^2}\sum_{i_j=1}^{C_j^2}\left(e_x^{i_j}\right)^2 + \frac{D_y^2}{6C_j^2}\sum_{i_j=1}^{C_j^2}\left(e_y^{i_j}\right)^2 \ .$$

Hence, Lemma-1 is proved.

APPENDIX-B: PROOF OF LEMMA-2

In a fully-spoofed scenario, the spoofing signal is strong enough to invade all the receivers in the ROI. Therefore, the double differential measurement in (9) degrades to:

$$\nabla\Delta\rho_{nm}^{ij} = \left(|S_f^i - p_f| - |S_f^i - p_f|\right) - \left(|S_f^j - p_f| - |S_f^j - p_f|\right) + \nabla\Delta\varepsilon_{nm}^{ij} \approx \nabla\Delta\varepsilon_{nm}^{ij}, \quad (B.1)$$

where $S_f^*$ and $p_f$ are the counterfeit satellite position and receiver position, respectively. Since the pseudorange measurements between different satellites and receivers are independent, $\nabla\Delta\rho_{nm}^{ij}$ follows a normal distribution:

$$\nabla\Delta\rho_{nm|f}^{ij} \sim \mathcal{N}(0, 4\sigma_\rho^2), \quad (B.2)$$

where $\sigma_\rho^2$ is the mean variance of the pseudorange errors of different satellites and receivers. Thus, the final distribution of the D²PS sample set will also be normal and its variance is readily obtained as $\sigma_{\rho|f}^2 = 4\sigma_\rho^2$. Hence, Lemma-2 is proofed.

In practice, the pseudorange error variance for each satellite is different according to the elevation, the signal SNR, and the multipath. However, since the D²SP sample set stands for the statistical distribution of the ensemble errors of different satellites and receivers, the usage of a mean variance $\sigma_\rho^2$ is reasonable.

APPENDIX-C: PROOF OF LEMMA-3

In a partially-spoofed scenario, the receivers in the monitoring area can be classified into two subsets: an authentic set denoted as $\mathcal{R}_a = \{p_m, m\forall(1-\alpha)M\}$ and a spoofed set denoted as $\mathcal{R}_f = \{p_l = p_f, l\forall\alpha M\}$. According to the generation process of D²SP sample set, there will be three different mutual differential operations. The first one comes from the double differential operations between receivers in the authentic set $\mathcal{R}_a$ following the normal distribution given in (12).

However, the spread area of the authentic set receivers is smaller than the entire monitoring area because some area are spoofed. Assuming that those receivers are uniformly spread in the non-spoofed area: $x_m \sim \mathcal{U}(a_1^a, a_2^a)$ and $y_m \sim \mathcal{U}(b_1^a, b_2^a)$, where $a_1 \leq a_1^a < a_2^a \leq a_2$ and $b_1 \leq b_1^a < b_2^a \leq b_2$. Then, we can define $D_x^a = a_2^a - a_1^a$ and $D_y^a = b_2^a - b_1^a$. According to the derivation of Lemma-1, the differential measurements for the authentic set $\mathcal{R}_a$ will have a normal distribution $\mathcal{N}_1(0, \sigma_{\nabla\Delta\rho_{mn}|a}^2)$ with the variance given by (14.c).

The second type of the measurement data comes from the operations in the spoofed set $\mathcal{R}_f$, whose distribution function is the same as (13), i.e., $\mathcal{N}_2(0, 4\sigma_\rho^2)$. Since the data is only affected by the measurement errors, its variance will be the same as that in the fully-spoofed scenario.

The third type comes from the differential operations between the spoofed set $\mathcal{R}_f$ and the authentic set $\mathcal{R}_a$. According to (9), for this type of differential operation, it has

$$\nabla\Delta\rho_{lm}^{ij} = \left(|S_f^i - p_f| - |S^i - p_m|\right) - \left(|S_f^j - p_f| - |S^j - p_m|\right) + \nabla\Delta\varepsilon_{lm}^{ij}$$
$$\approx \left(e_f^i - e_f^j\right)^\mathrm{T}\underbrace{(p_f - p_m)}_{\Delta p_{lm}} + \nabla\Delta\varepsilon_{lm}^{ij} \ . \quad (C.1)$$

In (C.1), it is assumed that the counterfeit position $p_f$ deviates from the true position by a few hundred meters or a few kilometers. This assumption is reasonably because a spoofer typically tries to deceive a receiver to a seemingly reasonable position. If the counterfeit position is too far from the actual position, such as hundreds of kilometers, it will be easily detected by simple information crosschecks. Furthermore, it also assumes in (C.1) that the counterfeit satellite position $S_f^i$ is approximately equal to the real satellite position $S^i$. This assumption applies to the meaconer or receiver-repeater. Even in the case of a simulator-based spoofer, it will also try to simulate actual satellite positions in order to make the spoofing signal appear more realistic.

Considering there are two opposite differential terms for each pair of receivers, i.e., $\Delta p_{lm} = -\Delta p_{ml}$, it can be deduced that the distributions of the position differences are piecewise uniform:

$$p(\Delta x_{lm}) \sim \begin{cases} 0.5\mathcal{U}(a_1^a - x_f, a_2^a - x_f), & a_1^a - x_f < \Delta x_{lm} < a_2^a - x_f \\ 0.5\mathcal{U}(x_f - a_2^a, x_f - a_1^a), & x_f - a_2^a < \Delta x_{lm} < x_f - a_1^a \\ 0, & \text{otherwise} \end{cases}$$
$$p(\Delta y_{lm}) \sim \begin{cases} 0.5\mathcal{U}(b_1^a - y_f, b_2^a - y_f), & b_1^a - y_f < \Delta y_{lm} < b_2^a - y_f \\ 0.5\mathcal{U}(y_f - b_2^a, y_f - b_1^a), & y_f - b_2^a < \Delta y_{lm} < y_f - b_1^a \\ 0, & \text{otherwise} \end{cases} \quad (C.2)$$

It can be seen that $E[\Delta x_{lm}] = 0$ and $E[\Delta y_{lm}] = 0$, and their variances are

$$\sigma_{\Delta x_{lm}}^2 = \frac{1}{3}[(x_f - a_1^a)^2 + (x_f - a_2^a)^2 + (x_f - a_1^a)(x_f - a_2^a)]$$
$$\sigma_{\Delta y_{lm}}^2 = \frac{1}{3}[(y_f - b_1^a)^2 + (y_f - b_2^a)^2 + (y_f - b_1^a)(y_f - b_2^a)] \quad (C.3)$$

Then, the distribution of $\nabla\Delta\rho_{lm}^{ij}$ can be approximated as

$$p(\nabla\Delta\rho_{lm}^{ij}) \sim e_x^{ij} p(\Delta x_{lm}) + e_y^{ij} p(\Delta y_{lm}) \ . \quad (C.4)$$

After averaging over all combinations of satellite pairs, the final probability density function of the differential measurements between spoofed and authentic sets will be the linear sum of $C_j^2$ independent random values whose distributions follow (C.4). According to the large number theorem, its probability density can be approximated by a normal distribution:

$$p(\nabla\Delta\rho_{lm}) \sim \mathcal{N}_3(0, \sigma_{\nabla\Delta\rho_{lm}|s'}^2), \quad (C.5)$$

where

$$\sigma_{\nabla\Delta\rho_{lm}|s'}^2 = \frac{\sigma_{\Delta x_{lm}}^2 \sum_{i_j=1}^{C_j^2}\left(e_x^{i_j}\right)^2 + \sigma_{\Delta y_{lm}}^2 \sum_{i_j=1}^{C_j^2}\left(e_y^{i_j}\right)^2}{C_j^2} \ . \quad (C.6)$$

Finally, we will see that the D²PS distribution will be the linear sum of three normally-distributed random variables, $\mathcal{N}_1\left(0, \sigma^2_{\nabla\Delta\rho_{mn}|a'}\right)$, $\mathcal{N}_2(0, 4\sigma^2_\rho)$, and $\mathcal{N}_3\left(0, \sigma^2_{\nabla\Delta\rho_{fm}|s'}\right)$:

$$p(\nabla\Delta\rho_{m_n})_{|af} = \frac{((1-\alpha)M-1)(1-\alpha)M}{(M-1)M}\mathcal{N}_1 + \frac{(\alpha M-1)\alpha M}{(M-1)M}\mathcal{N}_2 + \frac{2(1-\alpha)M\alpha M}{(M-1)M}\mathcal{N}_3 \ . \quad (C.7)$$

Since $M \gg 1$, (C.7) can be simplified as

$$p(\nabla\Delta\rho_{m_n})_{|af} \approx (1-\alpha)^2 \mathcal{N}_1 + \alpha^2 \mathcal{N}_2 + 2\alpha(1-\alpha)\mathcal{N}_3 \ . \quad (C.8)$$

It is known that $p(\nabla\Delta\rho_{m_n})_{|af}$ will be a normal distribution with a zero mean as well. It is noticed that $\mathcal{N}_1$ and $\mathcal{N}_2$ are uncorrelated, whereas $\mathcal{N}_1$ and $\mathcal{N}_3$, $\mathcal{N}_2$ and $\mathcal{N}_3$ are correlated and their correlation coefficients are both $2\alpha(1-\alpha)$. Hence, the variance of $\nabla\Delta\rho_{m_n}$ is derived as

$$\sigma^2_{\rho|af} = \sigma^2_{\rho|af} = (1-\alpha)^4 \sigma^2_{\nabla\Delta\rho_{mn}|a'} + 4\alpha^4 \sigma^2_\rho + 4\alpha^2(1-\alpha)^2 \sigma^2_{\nabla\Delta\rho_{lm}|s'} + 2\alpha(1-\alpha)\left(\sigma_{\nabla\Delta\rho_{mn}|a'} + 2\sigma_\rho\right)\sigma_{\nabla\Delta\rho_{lm}|s'} \ .$$

This concludes the Lemma-3 proof.